\documentstyle[aps,preprint,tighten]{revtex} 
\newcommand{\be}{\begin{equation}} 
\newcommand{\ee}{\end{equation}} 
\newcommand{\beq}{\begin{eqnarray}} 
\newcommand{\eeq}{\end{eqnarray}}

\newcommand{\tE}{\lefteqn{\smash{\mathop{\vphantom{<}}\limits^{\;\sim}}}E} 
\newcommand{\tP}{\lefteqn{\smash{\mathop{\vphantom{<}}\limits^{\;\sim}}}P} 
\newcommand{\tQ}{\lefteqn{\smash{\mathop{\vphantom{<}}\limits^{\;\sim}}}Q} 
 
\newcommand{\Pt}{\lefteqn{\smash{\mathop{\vphantom{\Bigl(}}\limits_{\sim} 
\atop \ }}P} 
\newcommand{\Qt}{\lefteqn{\smash{\mathop{\vphantom{\Bigl(}}\limits_{\sim} 
\atop \ }}Q}

\newcommand{\R}{R} 
 
\newcommand{\SA}{{\cal A}} 
\newcommand{\SSA}{{\bf A}}

\newcommand{\im}{\beta} 
\newcommand{\Rb}{{\rm \bf R}} 
\newcommand{\Cb}{{\rm \bf C}} 
\newcommand{\Nat}{{\rm \bf N}} 
 
\newcommand{\G}{{\cal G}} 
 
\newcommand{\CH}{{\cal H}} 

\newcommand{\CX}{{\cal X}}
\newcommand{\CXG}{{\cal X}_0}
\newcommand{\CXX}{{\tilde {\cal X}}}

\newcommand{\pr}{I_{(\chi)}^{(j)}} 
\newcommand{\Ppr}[2]{I_{(#2)}^{(#1)}} 

\newcommand{\tq}{{\tilde q}}
\newcommand{\tp}{{\tilde p}}
\newcommand{\tj}{{\tilde j}}
\newcommand{\ta}{{\tilde a}}

\newcommand{\hH}{{\bar h}} 
\newcommand{\gH}{{\bar g}}
\newcommand{\gG}{{{\hat g}_{\chi}}} 
\newcommand{\gGG}[1]{{{\hat g}_{\chi^{#1}}}} 

\newcommand{\UR}[1]{R(#1)}
\newcommand{\URm}[1]{R^{-1}(#1)}
\newcommand{\ve}{e} 
\newcommand{\fase}{\varphi}
\newcommand{\repr}{\rho}
\newcommand{\rep}{{\tilde \rho}}
\newcommand{\eps}{\varepsilon}
\newcommand{\ro}{\varrho} 
\newcommand{\p}{\partial}
\newcommand{\ph}{{\rm ph}}
\newcommand{\nn}{{m}}
\newcommand{\ver}{{v}} 
 
\newcommand{\IQ}{{I_{\smash{(Q)}}}}

\begin{document} 
% 
%%%%%%%%%%%%%%%%%%%%%%%%%%%%%%%%%%%%%%%%%%%%%%%%%%%%%%%%%%%%%%%%%%%%%% 
% 
\title{ 
{\Large \bf 
Hilbert space structure of covariant loop quantum gravity 
} 
} 
 
\author{ 
Sergei Alexandrov\thanks{e.mail: alexand@spht.saclay.cea.fr.  
Also at V.A.~Fock Department of Theoretical Physics, St.~Petersburg 
University, Russia} 
}  
 
\address{ 
Service de Physique Th\'eorique, C.E.A. - Saclay, 91191 Gif-sur-Yvette 
CEDEX, France\\ 
Laboratoire de Physique Th\'eorique de l'\'Ecole Normale 
Sup\'erieure, 24 rue Lhomond, 75231 Paris Cedex 05, France}

\maketitle 
 
\begin{abstract} 
We investigate the Hilbert space in the Lorentz covariant approach
to loop quantum gravity. We restrict ourselves to 
the space where all area operators are simultaneously diagonalizable,
assuming that it exists.
In this sector quantum states are realized by a generalization of
spin network states based on
Lorentz Wilson lines projected on irreducible representations of an SO(3)
subgroup. 
The problem of infinite dimensionality of the unitary Lorentz
representations is absent due to this projection. 
Nevertheless, the projection preserves the Lorentz covariance of 
the Wilson lines so that the symmetry is not broken.
Under certain conditions the states can be thought as functions on
a homogeneous space. We define the inner product as an integral
over this space. 
With respect to this inner product the spin networks form an
orthonormal basis in the investigated sector.
We argue that it is the only relevant part of a larger state space
arising in the approach.
The problem of the noncommutativity of the Lorentz
connection is solved by restriction to the simple representations.
The resulting structure shows similarities with 
the spin foam approach.
\end{abstract}

% 
%--------------------------------------------------------------------- 
\section{Introduction} 
%--------------------------------------------------------------------- 
% 

The loop approach to quantum gravity is a wide program of quantization
of general relativity (for review, see \cite{Rov-dif}). 
For last years it has achieved a large
progress in different directions. However, all this time it was
supplied with a number of problems which were little dark stains
on the nice picture drawn by loop quantum gravity.
Such problems as appearance of a nonphysical parameter
in the spectra of geometrical operators (Immirzi parameter problem
\cite{Imir,Rov-Tim}),
coincidence of the black hole entropy with the
quasiclassical result only up to a numerical factor \cite{entropy},
absence of an explicit relation with covariant formalism were considered as 
temporary difficulties which can not influence on the structure
of the theory and its basic conclusions.

But the situation turned out to be more complicated.
It has been shown that, actually, these problems are a manifestation
of deep problems of the chosen formulation itself. Namely,
it breaks the classical diffeomorphism invariance at the quantum level
\cite{SAcon}.\footnote{The first sign of this breaking has been given
by J.Samuel \cite{Sam}.} This happens due to a partial fixation of the gauge 
freedom reducing the gauge group from the Lorentz
one to SU(2). As it is well known, it is not allowed before quantization. 
Therefore, there arises a question about
the correctness of the whole program of loop gravity.
However, the formulation used so far to carry out the quantization
is only a particular realization of the ideas which the loop
approach relies on. So there may be a better formulation
which is free of the problems of the standard one. 

Recently, such a formulation, the so called covariant loop gravity,
was proposed \cite{SA}. Its main feature is that it possesses an
explicit Lorentz invariance and avoids any gauge fixing. 
In its framework the first results have been
already obtained. In particular, the area spectrum has been 
derived \cite{AV,SAcon} and the result
differs essentially from the one obtained in the approach with
the SU(2) gauge group \cite{area,ALarea}. 
It does not depend on the Immirzi parameter, strictly positive
and Lorentz invariant. Also the path integral has been shown to be
independent of the Immirzi parameter \cite{SA}.   
Among other pleasant features of this approach we mention
the polynomiality of the Hamiltonian constraint.

However, a quantization of the new formulation
on the level of Hilbert space was still lacking. 
There are several reasons for that.
The main reason is the noncompactness of the gauge group and 
all problems coming with it. To illustrate them it is enough 
to give one example: even Wilson loops of a Lorentz connection 
in an unitary representation are not well defined 
since one has to trace over an infinite dimensional space.
Another reason is the noncommutativity of the Lorentz connection
appearing as a canonical variable in the covariant approach \cite{SA}.

In this paper we are going to attack these problems altogether.
Since we do not know how the state space of general relativity is
embedded into the space of Lorentz Wilson loops,
our strategy will be, in a sense, opposite. At first, we assume that 
the Hilbert space can be constructed from the states
which are eigenstates of all area operators in a direct analogy
with the SU(2) case. We derive 
the structure of such states from the known result for the area
spectrum. It is done in Secs. II and III, 
where we end up with well defined Lorentz spin network states.
Then in Sec. IV we show that it is possible to introduce the Hilbert
space structure on the resulting space. We argue that 
it is sufficient to describe
all gravitational degrees of freedom. 
However, we are not able to give a strict proof of this assumption.
Also we obtain some restrictions on
the representations to be used and, surprisingly, they give
a solution of the noncommutativity problem.
Besides, we find a lot of correspondences with spin foam models 
\cite{sfLor} what indicates that we are not far from construction
of a solid bridge between these two approaches to quantum gravity.  
This and other issues are discussed in Sec. V. In two appendices one
can find some basic results on the covariant canonical formulation and
representations of the Lorentz group.

One should say that the paper does not claim for a high level of 
mathematical rigorousness. 
For example, we not always indicate explicitly in what mathematical sense some 
equations should be understood, to what spaces some quantities belong.
Rather our aim is to present the main ideas how
the Hilbert space can look like. Nevertheless, we try to show
that the emerging picture is self-consistent and very interesting
in many respects.

We use the following notations for indices. 
The indices 
$i,j,\dots$ from the middle of the alphabet label the space coordinates.
The latin indices $a,b,\dots$ from the beginning of the alphabet 
are the $so(3)$ indices, whereas 
the capital letters $X,Y,\dots$ from the end of the alphabet are 
the $so(3,1)$ indices.

% 
%------------------------------------------------ 
\section{Projected Wilson lines} 
%------------------------------------------------ 
% 

In \cite{AV,SAcon} it has been found that to be an eigenstate
of a quantum area operator the Wilson line  
\be 
U_{\alpha}[\SA]={\cal P}\exp\left(\int_a^b dx^i \SA_i^X T_X\right), 
\label{wl}
\ee 
should be defined by the Lorentz connection $\SA_i^X$ given by 
Eq. (\ref{spcon}). 
The corresponding area spectrum is given by eigenvalues of two 
Casimir operators 
\beq 
& {\cal S}= 8\pi\hbar G \sqrt{C(so(3)) -C_1(so(3,1))}, & \label{as} \\
& C_1(so(3,1))=g^{XY}T_XT_Y, & \label{L1Cas} \\
& C(so(3))=I_{(Q)}^{XY}T_XT_Y. & \label{SCas}
\eeq 
 
In this section we are going to give a more detailed construction 
of the Wilson lines diagonalizing the area operators. 
The problem is that the Wilson line (\ref{wl}) 
is an element of the Lorentz group, whereas the spectrum (\ref{as}) 
contains the Casimir of its SO(3) subgroup. This means that 
to be an eigenstate of the area, the Wilson line must be in definite 
irreducible representations of both SO(3,1) and SO(3). 
One should emphasize that we do not require it to be  
an element of the SO(3) subgroup. This requirement means only 
that the generators of this subgroup are in the given representation, 
whereas the Wilson line itself is an operator acting in the space 
of this representation. Nevertheless, in the following we shall use 
the shorten terminology. 
 
The Lorentz group contains a lot of possible embeddings of SO(3). 
Which subgroup should 
be considered is defined by the value of the field $\chi$ (see Appendix A). 
This dependence comes from the projector $I_{(Q)}^{XY}$ entering 
the Casimir $C(so(3))$. 
Therefore, if we want that the Wilson line gives a definite 
area for any surface, we must require that 
being cut at any point of the curve $\alpha$, 
it should be in the same representation of the SO(3) subgroup 
defined by the value of $\chi$ at this point. 
Thus, we arrive at the picture where the subgroup 
to which representation the Wilson line should be restricted 
is "rotated" along the line. 
 
How can this be realized? First of all, to pick out a part of the Wilson 
line, which is in a definite representation of the SO(3) subgroup, 
one can act to the ends of the line by  
the corresponding projector on this representation.\footnote{
This approach has been suggested by Carlo Rovelli.} 
The projector is given by the so-called {\it projective operator} \cite{BR}:
\be 
\pr=d_j\int_{SO(3)^{\chi}} d\mu(h) \overline{\chi^{j}(h)}\UR{h}, 
\label{proj}
\ee 
where $\chi^{j}(h)=\sum\limits_a D^j_{aa}(h)$ 
is the character of the representation
$j$, $d_j=2j+1$ is its dimension, $\UR{h}$ is a unitary representation of the
element $h$, and the integral is over the SO(3) subgroup defined by $\chi$
with the Haar measure. 
It is easy to obtain the following properties of this projector:
\beq
&\Ppr{j_1}{\chi}\Ppr{j_2}{\chi}=\delta_{j_1j_2}\Ppr{j_1}{\chi},& \\
&\UR{h} \pr=\pr \UR{h}. & \label{com-proj}
\eeq
Besides, the crucial fact for our construction 
is that it transforms in the 
covariant way under the Lorentz transformations. 
Indeed, we have\footnote{Throughout this paper, superscripts like $g$ or $h$
mean the corresponding (local) group transformations.}
\be 
\Ppr{j}{\chi^g}=d_j\int_{gSO(3)^{\chi}g^{-1}} d\mu(h) 
\overline{\chi^{j}(h)}\UR{h}
=d_j\int_{SO(3)^{\chi}} d\mu(h) 
\overline{\chi^{j}(h)}\UR{ghg^{-1}}=\UR{g} \pr \URm{g}. 
\label{proj-trans}
\ee
 
To introduce the picture of the "rotated" subgroup, this projector should 
be inserted in each point of the line. 
A rigorous definition of such a Wilson line can be given by use 
of a partition of the line into small pieces 
$\alpha=\bigcup_{n=1}^N \alpha_n$. In this case it is defined as a limit  
of infinitely many insertions of the projector (\ref{proj}). Let the
Wilson line (\ref{wl}) is in an irreducible representation 
$\lambda=(l_0,l_1)$ of the Lorentz group (see Appendix B). 
Then we define the so-called {\it projected Wilson line}: 
\be 
U^{(\lambda,j)}_{\alpha}[\SA,\chi]= \lim\limits_{N \rightarrow \infty} 
{\cal P}\left\{ \prod\limits_{n=1}^{N} 
\pr(a_{n+1}) U_{\alpha_n}[\SA] \pr(a_{n}) \right\}. 
\label{WLp}
\ee 
 
The projected Wilson lines are operators acting in  
finite dimensional spaces 
of irreducible representations of the SO(3) group, despite they 
are defined by the Lorentz connection $\SA_i^X$. 
(It is worth to note that this finiteness allows the Wilson loops to 
be well defined in the sense that they produce finite numbers. 
Otherwise, we would have to trace an infinite dimensional matrix 
and at least in the case of vanishing connection the answer would be 
definitely divergent.) 
Nevertheless, they arise from Lorentz group elements and
it is important that they would transform in the standard covariant way 
under the local Lorentz transformations: 
\be 
U^{(\lambda,j)}_{\alpha}[\SA,\chi] \longrightarrow 
R_{\lambda}(g(b)) U^{(\lambda,j)}_{\alpha}[\SA,\chi] R_{\lambda}^{-1}(g(a)), 
\label{tran-wl}
\ee 
where $R_{\lambda}(g)$ is a Lorentz group element 
in the representation $\lambda$ carried by the line.  
That this is indeed true can be easily seen from the property 
(\ref{proj-trans}).
This means that the Lorentz invariance is not broken 
by the projection. 
 
As we pointed out, in a general case the projected Wilson line  
does not belong to neither SO(3) nor SO(3,1).  
This fact makes the construction quite 
complicated since, for example, 
the usual definition of the cylindrical functions and construction
of the inner product \cite{cyl}
generalized to the case of the Lorentz group would not work --- 
the projected Wilson lines are not maps to any group.  
However, there is a way to avoid at least some of these problems. 
For this, note the following remarkable fact. 
Let us consider the Lorentz generators  
in an irreducible representation. Their matrix elements are given in Eqs. 
(\ref{gauss-rep}) and (\ref{boosts-rep}). Then project them on the  
subspace $\CH_j=\{\xi_{j,m}\}_{m=-l}^{l}$  
of an irreducible representation of the SO(3) subgroup  
which is chosen to be the canonically embedded one.  
To make this projection it is enough to restrict ourselves to  
the matrix elements like $<\xi_{j,m}|T_X|\xi_{j,m'}>$.   
Doing so we obtain that the projected generators satisfy   
\be  
\Ppr{j}{0}F_a\Ppr{j}{0}=\beta_{(j)} \Ppr{j}{0}H_a\Ppr{j}{0}. \label{prgen}
\ee  
This means that in the given representation 
the projected boost generators $F_a$ form the $su(2)$ algebra as $H_a$ do!  
   
Of course, it is not exactly what we want. We are interested in the  
group elements which are given by the Wilson lines rather  
than in generators. {\it A priori} the projected   
boosts do not form the SU(2) group or any other.  
Besides, there is a problem that in general we should project to   
representation spaces of different SO(3) subgroups at different ends  
of a line. Whether the resulting objects will form a group is  
a question.  
  
Nevertheless, we can overcome these obstacles in the following way.  
First of all, let us fix a gauge taking $\chi=0$.  
This always can be done  due to the transformation law (\ref{tran-wl}) 
at the expense of Lorentz group elements at the ends of the line: 
\be 
U^{(\lambda,j)}_{\alpha}[\SA,\chi] = 
R_{\lambda}^{-1}(\gG(b)) U^{(\lambda,j)}_{\alpha}[\SA^{\gG},0] 
R_{\lambda}(\gG(a)), 
\label{wl-gfix}
\ee  
where $\gG$ is the local Lorentz transformation sending $\chi$ to $0$.
Then all projectors become constant along all Wilson lines and this is the  
canonically embedded SO(3) which the Lorentz Wilson lines are projected on.  
  
Let us proceed with the definition (\ref{WLp}). 
For a sufficiently fine partition we can write  
\beq  
U^{(\lambda,j)}_{\alpha}[\SA,\chi]&=& {\cal P}\left\{ \prod\limits_n  
\pr(a_{n+1}) \left(1+\int_{\alpha_n}dx^i\SA_i^X T_X\right) \pr(a_{n})  
\right\} \\  
&=&  
{\cal P}\left\{ \prod\limits_n  
\pr(a_{n+1}) \left(1+\int_{\alpha_n}dx^i\pr(x)\SA_i^X T_X \pr(x)   
\right) \pr(a_{n}) \right\} \\  
&=& {\cal P}\left\{ \prod\limits_n  
\pr(a_{n+1}) U_{\alpha_n}[\pr\SA\pr] \pr(a_{n}) \right\}.  
\label{WLp-su2}
\eeq  
Due to the property (\ref{prgen}) in the time gauge we obtain  
\be  
U^{(\lambda,j)}_{\alpha}[\SA,0]= 
\iota_{\lambda}\left( U_{\alpha}[\SSA^{(j)}]\right), 
\label{WL-W2}
\ee  
where $\iota_{\lambda}$ denotes the embedding of an operator 
in a representation
of SU(2) into the representation $\lambda$ of SO(3,1) and 
$\SSA^{(j)}$ is the $su(2)$ connection given by  
\be  
\SSA_i^{(j)a}=\frac12 {\eps^a}_{bc}\omega_i^{bc}-\beta_{(j)} 
\omega_i^{0a}+O(\G_X).  
\label{conAB}
\ee  
As a result, the projected Wilson line turns out to be an element of the  
SU(2) group in the representation with the spin $j$. It is given by  
the ordinary SU(2) Wilson line with the connection coinciding   
(on the surface of the Gauss constraint) with  
the Ashtekar-Barbero connection \cite{barb} in which the Immirzi parameter   
is defined by the Lorentz and SU(2) representations of the initial  
Wilson line  
\be  
\beta_{(j)}=\frac{n\rho}{j(j+1)} \qquad (principal\ series).  
\label{param}
\ee  
  
However, one should be care using the representation (\ref{WL-W2}).  
It gives the right value of the projected Wilson line itself, 
but it can not be used in calculations involving commutators.  
For example, it contradicts to the area   
spectrum we started with. Indeed, instead of   
\be  
{\cal S}\sim \hbar  
\sqrt{ j(j+1) -n^2+ \rho^2+1},  
\label{areaspL}
\ee   
the Wilson line (\ref{WL-W2}) gives  
\be  
{\cal S}\sim \hbar |\beta_{(j)}| \sqrt{ j(j+1)}=  
\frac{\hbar n|\rho|}{\sqrt{j(j+1)}}.  
\label{areaspSU2}
\ee   
The reason for the discrepancy is that the projectors on SO(3)  
representations should be inserted only between operators in different  
points as the expression (\ref{WLp}) tells us.   
On the other hand, in the calculation of the area spectrum   
a quadratic expression in generators in the same point appears.  
In the Wilson line (\ref{WL-W2}) they are already SU(2) generators,  
whereas the correct way is to project only their quadratic combination.  
As a result, we have to conclude that the initial representation (\ref{WLp}), 
but not the simplified one (\ref{WL-W2}), should be used in this case.  

Actually, this means that the limit in Eq. (\ref{WLp}) does not 
commute with action of quantum operators due to their
distributional nature.
It can be illustrated as follows.
At finite, but sufficiently fine partition the projected 
Wilson line (\ref{WLp}) can be represented as 
the SU(2) Wilson line (\ref{WL-W2}) plus 
corrections of order $1/N$ and higher, where N is the degree of the partition.
Since $[\SA(x),\tP(y)]\sim \delta(x-y)$ all corrections of order less
than the number of commutators involved contribute to the result
and can not be neglected.

  Nevertheless, the important conclusion is that  
the projected Wilson lines in calculations without quantum commutators, 
up to Lorentz transformations at the ends 
arising due to a nonvanishing field $\chi$, can be considered as
SU(2) group elements in the representations which the Wilson lines 
are projected on.

% 
%--------------------------------------------------------------------- 
\section{Spin networks} 
%--------------------------------------------------------------------- 
% 

Let us neglect, for a moment, the noncommutativity 
of the connection used. Then one can construct natural spin network states 
from the Wilson lines described in the previous section. To this end,  
it is enough to associate them to the links of a graph $\Gamma$ and contract 
with interwiners of the Lorentz group at the nodes: 
\be 
\Psi_S(\SA,\chi)=\mathop{\otimes}\limits_{{\rm links \ } \gamma_i \in \Gamma} 
U^{(\lambda_i,j_i)}_{\gamma_i}[\SA,\chi]\cdot 
\mathop{\otimes}\limits_{{\rm nodes} \ \ver\in \Gamma}N_{\ver}. \label{spnetA}
\ee 
The subscript $S$ denotes the collection of the graph, SO(3,1) and SO(3)
representations assigned to the links, and interwiners at the
nodes: $S=(\Gamma,\{\lambda_i\},\{j_i\},\{N_{\ver}\})$.
The interwiners $N_{\ver}$ should 
be elements of the tensor product of representation spaces 
assigned to the links meeting at the node: 
\be 
\CH_{\ver}=\CH_{\lambda_1}\otimes \cdots \otimes \CH_{\lambda_k}. 
\label{Hp}
\ee 
(To get a Lorentz invariant spin network one should take interwiners 
to be invariant tensors only, i.e. 
lying in the trivial representations entering the decomposition 
of the product of representations (\ref{Hp}).) 
Due to the projectors at the ends of the Wilson lines, each index 
of the interwiners is projected to the corresponding 
subspace of an irreducible representation of SO(3). 
Thus, the problem of infinite dimensionality of the unitary
representations of the Lorentz group is avoided, since we trace,
actually, over finite dimensional spaces.
We emphasize that due to the transformation law (\ref{tran-wl})
the Lorentz invariance is not broken.
 
Let us apply to the constructed spin network states the transformation 
(\ref{wl-gfix}). 
The result can be written in the following way:
\beq 
\Psi_S(\SA,\chi)&=&\mathop{\otimes}\limits_{{\rm links \ } 
\gamma_i \in \Gamma} 
R_{\lambda_i}^{-1}(\gG(\ver^{(f)}_i))
U^{(\lambda_i,j_i)}_{\gamma_i}[\SA^{\gG},0]
R_{\lambda_i}(\gG(\ver^{(i)}_i)) \cdot 
\mathop{\otimes}\limits_{{\rm nodes} \ \ver\in \Gamma} N_{\ver}
\nonumber \\
&=&\mathop{\otimes}\limits_{{\rm links \ } \gamma_i \in \Gamma} 
\iota_{\lambda_i}\left(h_{i}^{(j_i)}\right)\cdot 
\mathop{\otimes}\limits_{{\rm nodes} \ \ver\in \Gamma} 
N_{\ver}^{\gG^{-1}(\ver)}, 
\label{spnet-tr}
\eeq 
where\footnote{We omitted the superscript $\gG$ for $\SSA^{(j)}$.
Actually, the $su(2)$ connection is given by Eq. (\ref{conAB})
with $\omega_i^{\alpha\beta}$ transformed by the local Lorentz
transformation $\gG$.}
 $h_{i}^{(j_i)}= U_{\gamma_i}[\SSA^{(j)}]\in SU(2)$,
$\gG(\ver)\in SO(3,1)$ and $\ver^{(i)}_i,\ver^{(f)}_i$ are the initial 
and final points of the $i$th link. 
In the second line we used the property (\ref{WL-W2}).
We see that the effect of a nonvanishing $\chi$ reduces to the Lorentz 
transformations of the connection and interwiners. In particular, for
the gauge invariant spin networks the latter transformation is absent
due to the Lorentz invariance of $N_{\ver}$. 
In the next section we will argue that these spin networks form a basis of
the Hilbert space of quantum gravity in the loop approach.

% 
%--------------------------------------------------------------------- 
\section{Hilbert space structure} 
%--------------------------------------------------------------------- 
% 
 
As a base of the loop approach, we assume that the physical excitations 
of quantum space are concentrated on one-dimensional structures, which are, 
in a general case, graphs. Then it is natural to suppose 
that the Hilbert space is formed by spin network-like states.
A generalization of the usual spin networks was presented in the previous 
section. However, we should check whether they span the whole state space,
find which representations should be taken into account and construct
an inner product on the resulting space. 

In this section we address these issues. The idea is to
associate with each quantum state a function on a homogeneous space.
Then the whole state space is given by the space of such functions
subject to conditions to be specified. 
One can apply the harmonic analysis on homogeneous spaces to find a basis
in this space, and
the inner product is given by the integral over the homogeneous space.
Once the inner product is defined on the space of functions 
and the correspondence with quantum states is established, 
it induces the Hilbert structure on the quantum state space. We emphasize
that the functional space plays in this construction an auxiliary role,
and there are certain restrictions on its identification with 
the quantum state space.

The crucial point is the choice of the homogeneous space.
At first, we motivate the choice analysing
the kinematical space of general relativity. However, as we will see,
this analysis is insufficient to find the right space since 
the representation of quantum states by functions on such space
fails to be always correct. 
Nevertheless, a little extension of the chosen homogeneous
space allows to make the emerging picture consistent with the previous
analysis of the quantum area operator, although a strict proof that it is
sufficient is still lacking.  

\subsection{The state space}

First of all, let us discuss the degrees of freedom which give rise
to the state space of general relativity with the SO(3,1) gauge group.
Now we have 18-component Lorentz connection $\SA$. 
So we can expect that the quantum states can be represented 
as functions of this 
connection.\footnote{Actually, they can not be
functions of the connection because it is noncommutative. However,
we can choose a vacuum and act on it by the operators constructed 
from these functions. In this sense one can establish a correspondence
between the Hilbert space and the space of functions of the connection.} 
Being associated with a graph, the connection should give rise 
to SO(3,1) holonomies. 
Thus, the naive analysis leads to the realization of the state space
by functions on the $G=SO(3,1)$ group manifold.

However, this is a very simplified picture.
Let us look at the constructed spin networks (\ref{spnetA}). 
The first question which arises is
why can we write a wave function as a functional of both the 
connection and the field $\chi$? 
It turns out that it is possible because they commute 
and the connection $\SA$ has just three independent components less 
than the canonical one $A$ \cite{AV}. 
The missed components are encoded in the Gauss constraint and they are not 
taken to be "configuration variables".
Moreover, there are the second class constraints which
fix six components more of the connection. As a result, we end up
with two objects $\SA$ and $\chi$ having 9 and 3 
independent components correspondingly.
$\SA$ is naturally associated with the links of graphs. However, 
taking into account the number of independent components and
regarding Eq. (\ref{spnet-tr}), it gives rise more naturally
to holonomies of a 3-dimensional group $H$
which is, of course, expected to be SU(2). What is the role of $\chi$?
From the explicit form of the diffeomorphism constraint \cite{AVpath}, when
the second class constraints are solved, one can see that the field 
$\chi$ transforms as a scalar. Due to this, it is natural
to associate it with space points, i.e., with the nodes of graphs.
This is exactly the same conclusion 
what can be found from the result (\ref{spnet-tr}).  
Besides, since $\chi$ is related to boosts only, it can be considered 
as a coordinate on the homogeneous space $X=SO(3,1)/SO(3)$. 
Thus, one can expect that the Hilbert space is realized by functions
on $H$ associated with the links and $X$ associated with the nodes.
Therefore, given a graph $\Gamma$ with $n$ links and $\nn$ nodes,
we choose the corresponding homogeneous space to be 
$\CXX=[H]^n\times[X]^{\nn}$.

However, we are interested not in all functions on $\CXX$. 
We should put in the information about the structure of the underlying graph 
and the symmetry properties.
To do this, we impose an important requirement that the functional space 
to be considered
carries a unitary representation of the Lorentz group and
the gauge invariant sector is given by the functions independent of $X$.
It restricts us to a subspace of the full space $L^2(\CXX)$.
To use this requirement one should decide how the elements of $H$ and
$X$ transform under the action of $G$. The transformations can be found from
the representation (\ref{spnet-tr}), which relates the elements to $\SA$ and
$\chi$ whose transformation laws under the local Lorentz group $\G$
are known. 
In particular, we identify an element $x\in X$ of the homogeneous space with
$\gG^{-1}$ and choose it to be a pure boost.
Since $\chi=0$ is a stationary point under the action of $H$,
this identification is an isomorphism.
 
Let $g\in \G$  and $g\gG^{-1}=\gH\hH$ is the Cartan
decomposition \cite{BR}. 
($\hH \in H$ and $\gH$ is a pure boost.)
From the definition of $\gG$ it follows that 
the transformation of $x\in X$ is given by the boost leading to
$\chi^g$ from the gauge $\chi=0$. Thus we have 
\be 
x=\gG^{-1} \longrightarrow \gGG{g}^{-1}= \gH = gx\hH^{-1}.
\label{trans-x}
\ee
Using this result we obtain
\be
\SA^{\gG} \longrightarrow \left(\SA^g\right)^{\gGG{g}}=
\left(\SA^{\gG}\right)^{\hH}.
\ee
Therefore
\be
\iota_{\lambda} (h)=U_{\gamma}^{(\lambda,j)}[\SA^{\gG},0] \longrightarrow 
U_{\gamma}^{(\lambda,j)}\left[\left(\SA^{\gG}\right)^{\hH},0\right]=
R_{\lambda}(\hH(b)) \iota_{\lambda}(h) R_{\lambda}^{-1}(\hH(a)).
\ee
Since $R_{\lambda}(\hH)\iota_{\lambda}(h)=\iota_{\lambda}(\hH h)$ 
due to Eq. (\ref{com-proj}), the transformation law 
is\footnote{As it is seen from the derivation the law (\ref{trans-h})
is also valid for the nonprojected Wilson lines 
$U_{\gamma}\left[\SA^{\gG}\right]$.}
\be
h \longrightarrow \hH(b) h \hH^{-1}(a).
\label{trans-h}
\ee
It is just to cancel the compensating rotation $\hH$ from Eq. (\ref{trans-x}).

Given the transformation laws (\ref{trans-x}) and (\ref{trans-h}),
we can formulate our requirement. 
Let $f_{\Gamma}(h,x)$ be a square integrable function 
associated with the graph $\Gamma$, where $h\in [H]^n$ and $x\in [X]^{\nn}$.
Then the following representation should exist:
\be
  f_{\Gamma}(h,x)=
 \left(\prod\limits_{r=1}^{\nn}\int\limits d\ro(\lambda_r) 
 \sum\limits_{p_r}\right)  
{\check f}_{p_1 \dots p_{\nn}}^{\lambda_1\dots\lambda_{\nn}}
 e_{p_1\dots p_{\nn}}^{\lambda_1\dots\lambda_{\nn}}(h,x),
\label{genf}
\ee
where $d\ro(\lambda)$ is a measure on the the set of irreducible
representations of $G$,
$p=(j,a)$ labels a basis in a given representation
and the vectors 
$e(h,x)$ are so that for any $g\in [G]^{\nn}$ we have 
\be
 e_{p_1\dots p_{\nn}}^{\lambda_1,\dots,\lambda_{\nn}}
(\hH_{(f)} h \hH^{-1}_{(i)}, gx\hH^{-1} )=
\left(\prod\limits_{r=1}^{\nn} \sum\limits_{q_r} 
D_{p_rq_r}^{\lambda_r}(g_r) \right) 
 e_{q_1\dots q_{\nn}}^{\lambda_1\dots\lambda_{\nn}}(h,x),
\label{gentr}
\ee
where $D_{pq}^{\lambda}(g_r)$ are matrix elements of $g_r\in G$ in the 
representation $\lambda$ and the Cartan decomposition defining $\hH$ was used.
Eq. (\ref{gentr}) means that $\hH^{-1}$ arising at each node should 
be cancelled by the corresponding transformations of $h$'s associated with
the adjacent links. Eq. (\ref{genf}) is nothing but the decomposition of
a general representation in a direct integral of irreducible
representations. The functions of $L^2(\CXX)$ satisfying the described symmetry
requirement will give a generalization of the usual cylindrical functions.

Let us investigate the resulting space of all admissible functions 
related to the given graph $\Gamma$. 
We decompose a given $f_{\Gamma}(h,x)$ in irreducible representations to
find a basis in this space.
At the first step by the Peter-Weyl theorem 
it can be represented as a sum over
all unitary irreducible representations of $H$:
\be
 f_{\Gamma}(h,x)=\left(\prod\limits_{i=1}^n\sum\limits_{j_i, a_i,b_i} 
 D^{j_i}_{b_ia_i}(h_i) \right)
 {\breve f}^{j_1\dots j_n}_{a_1\dots a_n,b_1\dots b_n}(x).
\ee
At the second step we expand the coefficient functions using the theorem
given by the harmonic analysis on the homogeneous spaces \cite{BR}:
\be
{\breve f}^{j_1 \dots j_n}_{a_1\dots a_n,b_1\dots b_n}(x)=
 \left(\prod\limits_{r=1}^{\nn}\int\limits_{0}^{\infty} d\rep_r \, \rep_r^2
 \sum\limits_{p_r,q_r}  D^{(0,i\rep_r)}_{p_rq_r}(g_{x_r}) 
 e_{q_r}(\rep_r)\right)
 {\bar f}^{j_1\dots j_n, p_1\dots p_{\nn}}_{a_1\dots a_n,
b_1\dots b_n} (\rep_1,\dots,\rep_{\nn}),
\label{coef1}
\ee
where $g_x$ is a representative of $x$ in $G$ 
and $e_q(\rep)$ is a stationary vector of $H$: 
$D^{(0,i\rep)}_{pq}(h)e_q(\rep)=e_p(\rep)$. 
Notice, that only the so called {\it simple} representations of $G$
of the principle series with $n=0$ contribute to the expansion.
The coefficient functions $\bar f$ are restricted by the conditions
(\ref{genf}) and (\ref{gentr}).
The most general $\bar f$ satisfying
them can be given in terms of the Clebsch-Gordon coefficients
since they are only objects relating indices in different
representations.
The result is the following:
\beq
& {\bar f}^{j_1\dots j_n,p_1\dots p_{\nn}}_{a_1\dots a_n,b_1\dots b_n} 
(\rep_1,\dots,\rep_{\nn})
 =\Biggl(\prod\limits_{r=1}^{\nn}
\int_X d\mu(x_r) 
\overline{ D^{(0,i\rep_r)}_{p_rq_r}(g_{x_r}) e_{q_r}(\rep_r)} 
\times   \Biggr. & \nonumber \\
&\times \Biggl.
\int d\ro(\lambda_r)  \sum\limits_{\tp_r}
\sum\limits_{\tq_r=({\tj}_r,\ta_r)} 
D^{{\lambda}_r}_{\tp_r\tq_r}(g_{x_r})
<\tq_r|j_{r_k},a_{r_k};j_{r_l},b_{r_l}>_H  \Biggr)
 {\tilde f}^{j_1\dots j_n}_{\tp_1\dots\tp_{\nn}, \tj_i\dots\tj_{\nn}} 
({\lambda}_1,\dots,{\lambda}_{\nn}). &
\label{coef02}
\eeq  
In our notations ${r_k}$ and $r_l$ refer to all outgoing 
and incoming links for $r$th node correspondingly. $< \cdot|\cdots>_H$
denotes SU(2) interwiners for the tensor product of any number
of representations.
Substitution of the expression (\ref{coef02}) in Eq. (\ref{coef1}) gives
\beq
& f_{\Gamma}(h,x)= 
 \left(\prod\limits_{i=1}^n\sum\limits_{j_i, a_i,b_i} 
 D^{j_i}_{b_ia_i}(h_i) \right)
 \Biggl(\prod\limits_{r=1}^{\nn}\int d\ro(\lambda_r) 
 \sum\limits_{\tp_r,\tq_r}  D^{\lambda_r}_{\tp_r\tq_r}(g_{x_r}) 
\times \Biggr. & \nonumber \\
& \times \Biggl. 
<\tq_r| j_{r_k},a_{r_k}; j_{r_l},b_{r_l}>_H   \Biggr)
 {\tilde f}^{j_1\dots j_n}_{\tp_1\dots\tp_{\nn},\tj_i\dots\tj_{\nn}} 
(\lambda_1,\dots,\lambda_{\nn}). &  
 \label{coef03}
\eeq
The property (\ref{gentr}) follows from two facts that the Lorentz
group matrix elements of an element of the SO(3) subgroup do not
depend on the representation of the Lorentz group  
and they are diagonal with respect to the index $j$ (see
(\ref{gauss-rep})). Due to this 
\be
D^{\lambda}_{\tp\tq}(g_{gx\hH^{-1}})=\sum\limits_{p,q}
D^{\lambda}_{\tp p}(g)D^{\lambda}_{pq}(g_{x})
D^{\lambda}_{q\tq}(\hH^{-1})=
\sum\limits_{p}\sum\limits_{a}
D^{\lambda}_{\tp p}(g)D^{\lambda}_{pq}(g_{x})
D^{\tj}_{a\ta}(\hH^{-1}),
\ee
where in the last expression $q=(\tj,a)$, $\tq=(\tj,\ta)$.
$D^{\tj}_{a\ta}(\hH^{-1})$ can be brought through the
interwiner to act on the adjacent links and cancel
the result of their transformations.

To establish a correspondence between the functions (\ref{coef03}) 
and the spin networks (\ref{spnet-tr}),  
we associate to each link (or $h_i\in H$) 
a simple representation $(0,i\repr_i)$ of the Lorentz group. 
The restriction to the representations with $n=0$ is due to that 
each of them contains all representations of $H$, whereas any other
restricts to $j\ge n$. Therefore we can embed a function on $H$
without loss of any information only into a simple representation.
Besides, we change the SU(2) interwiners by their Lorentz 
counterparts. (We omit the corresponding subscript.) 
Then the function (\ref{coef03}) becomes
\beq
& f_{\Gamma}(h,x)=\left(\prod\limits_{i=1}^n\sum\limits_{j_i, a_i,b_i} 
 \iota_{(0,i\repr_i)}\left(D^{j_i}_{b_ia_i}(h_i)\right) 
\int\limits_{-\infty}^{\infty} d\repr_i \, \repr_i^2  \right)
 \Biggl(\prod\limits_{r=1}^{\nn} \int d\ro(\lambda_r) 
 \sum\limits_{\tp_r,\tq_r}  D^{\lambda_r}_{\tp_r\tq_r}(g_{x_r}) 
\times \Biggr. & \nonumber \\
& \times \Biggl. <\lambda_r, \tq_r| -\repr_{r_k}, (j_{r_k},a_{r_k});
\repr_{r_l}, (j_{r_l},b_{r_l})>   \Biggr)
 {\hat f}^{j_1\dots j_n}_{\tp_1\dots\tp_{\nn},
\tj_i\dots\tj_{\nn}} 
(\lambda_1,\dots,\lambda_{\nn}; \repr_1,\dots,\repr_n). &
\label{coef3}
\eeq
Indeed, it is the same as in Eq. (\ref{coef03}). 
It can be seen due to the following factorization property \cite{Cl-G}:
\be
 <\lambda, (\tj,\ta)| \repr_{1}, (j_1,a_1);
 \repr_2, (j_2,a_2)> = 
<\tj,\ta| j_1,a_1;j_2,a_2>_H F(\lambda, \tj| \repr_1,j_1; \repr_2, j_2). 
\label{factor-pr}
\ee
Due to this only $F$ and $\hat f$ depend on $\repr_i$ and we obtain
an expression for $\tilde f$ as an integral of $\hat f$ with the
functions $F$ over $\repr_i$.

If the coefficients $\hat f$ independent of the indices $\tj$,
in the result (\ref{coef3}) one can recognize an arbitrary 
linear combination of the states (\ref{spnet-tr}). 
What does the dependence of $\tj$ mean? In fact, the states (\ref{spnetA}) 
possess a larger invariance than we required from our functions.
They are explicitly invariant under simultaneous Lorentz
transformations of the Wilson lines and interwiners. In our terms
it means
\be 
h \longrightarrow g(b)hg^{-1}(a), \qquad 
x \longrightarrow xg^{-1}.
\label{exttrans}
\ee
The problem is that it takes away the arguments from their spaces and,
therefore, the requirement of the invariance under this transformation 
can not be formulated in terms of the functions on
$\CXX$ only. We do not know how to implement
it. Therefore, we simply postulate that only the functions
(\ref{coef3}) with $\hat f$ independent of $\tj$ should be considered.
Notice, that in the gauge invariant sector this problem does not arise,
since in this case $\tj$ takes only one value $\tj=0$. 

Thus, we obtain that linear combinations of 
the states (\ref{spnet-tr}) span all functions
$f_{\Gamma}(h,x)$ subject to the described conditions. 
However, from the above it is
clear that not all states (\ref{coef3}) differing
only by the Lorentz representations $\repr_i$ assigned to the links
are described by different functions. 
Most explicitly it can be seen for the graph consisting of one loop.
In this case the function (\ref{coef3}) does not depend on $\repr$ at all.
This means that in the described space the area operator can not be 
implemented as a self-adjoint operator.
Indeed, the area spectrum (\ref{areaspL}) ``feel'' 
the Lorentz representations and, as it is known,
eigenstates with different eigenvalues must be orthogonal.
But in our case the eigenstates differing only by $\repr_i$ would not be
orthogonal with respect to the inner product induced by 
the natural inner product on the space $L^2(\CXX)$
(see the next subsection).
Therefore, the representation of quantum states by the functions 
$f_{\Gamma}(h,x)$ is essentially incomplete. Actually, we have already 
seen this in the end of Sec. II, where it was argued that 
the change of the projected Wilson lines by SU(2) elements is valid 
until quantum commutators are involved in calculations.

We suggest a simple way to improve the situation. We argue that 
it is sufficient to associate with the links an additional 
variable, say $\fase_i$, to make the resulting picture self-consistent.  
The modification can be interpreted as we take into
account quantum effects lost after the change of the Wilson lines 
by the elements of $H$, which has been done in Eq. (\ref{WL-W2}). 
The new variable takes values in $\Rb$ and it distinguishes the states with
different $\repr_i$. Indeed, now our state space is the space of functions
$f_{\Gamma}(h,\fase,x)$ on the homogeneous space 
$\CX=[H\times \Rb]^n\times [X]^{\nn}$ subject to the previous conditions
(\ref{genf}) and (\ref{gentr}). 
We can expand its elements in irreducible representations as above. 
In this way we arrive at Eq. (\ref{coef03}), where the coefficients $\tilde f$
are functions of $\fase$. Therefore we should expand them in the ordinary 
Fourier integral what gives the additional factor
$ \int\limits_{-\infty}^{\infty} d\repr_i \ e^{i\fase_i\repr_i}$
for each link. Besides, the coefficients become functions of $\repr_i$.
Then we can redefine them by the functions $F$ from
Eq. (\ref{factor-pr})
and take to be independent of the indices $\tj$ as discussed above.
As a result, we arrive at the following representation:
\be 
f_{\Gamma}(h,\fase,x)=\left(\prod\limits_{i=1}^n\sum\limits_{j_i} 
\int\limits_{-\infty}^{\infty} d\repr_i \, \repr_i^2  \right)
 \left(\prod\limits_{r=1}^{\nn} \int d\ro(\lambda_r) 
 \sum\limits_{p_r}    \right)
 {\hat f}^{j}_{p}(\lambda; \repr)
\ve^{j}_{p}(\lambda; \repr),
\label{coef4}
\ee
where
\be
\ve^{j}_{p}(\lambda; \repr)=
\left(\prod\limits_{i=1}^n\sum\limits_{a_i,b_i} 
 D^{j_i}_{b_ia_i}(h_i) e^{i\fase_i\repr_i} \right)
 \prod\limits_{r=1}^{\nn} \sum\limits_{q_r}  
D^{\lambda_r}_{p_rq_r}(g_{x_r})
\frac{<\lambda_r, q_r| -\repr_{r_k}, (j_{r_k},a_{r_k});
 \repr_{r_l}, (j_{r_l},b_{r_l})>}{ 
N_r(\lambda_r; \{ \repr_{r_k}, j_{r_k}\},\{ \repr_{r_l}, j_{r_l}\})}  
\label{basis1}
\ee
and we used shorten notations for indices and arguments.
The normalization factors $N_r$ will be found below.

The functions (\ref{coef4}) describe our state space
and the vectors (\ref{basis1}) form a basis in it.
The main result of this subsection is the information
about representations to be taken into account.
From Eq. (\ref{coef4}) we conclude:
1) Only the simple representations of type $(0,i\repr)$ should be
associated with links. 
2) Since only representations with $j\in \Nat$ enter 
the decomposition of the representations $(0,i\repr_i)$,
in Eq. (\ref{coef4}) we actually sum over integer $j$'s only.
Therefore, $H=SO(3)$ rather than SU(2). Actually, it is natural
since $H$ must be a subgroup of SO(3,1). 
3) Similarly, only $\lambda=(l_0,l_1)$, $l_0\in \Nat$ appear in the states
what follows from the properties of the Clebsch-Gordon coefficients.
The only restriction on $l_1$ is that it corresponds to the principle
series of representations, i.e., $l_1=i\rep$. 

The fact that it is sufficient to consider the Wilson lines in
the simple representations $(0,i\repr)$ only has a very important consequence.
For such representations the effective Immirzi parameter (\ref{param})
$\beta_{(j)}=0$. Due to this two projected Wilson lines (\ref{WLp})
commute with each other
despite of the noncommutativity of the connection!
Indeed, their commutator gives rise to
$T_X [\SA_i^X,\SA_j^Y]T_Y$. The generators act in different
representation spaces and so they can be projected.
After this it is enough to obtain that the commutator
of $so(3)$ components of the connection
$[I_{(Q)Z}^X\SA_i^{Z},\SA_j^{W}I_{(Q)W}^Y]$ vanishes.
This can be shown by tedious but direct
calculations from the result (\ref{AAcom}).
As a result, the problem of the noncommutativity
disappears for the constructed states and the spin
networks (\ref{spnetA}) are defined unambiguously.

\subsection{The inner product}

On the described above state space one can define a
natural inner  product. Since our sates are realized by functions
on $H\times \Rb$ and $X=SO(3,1)/SO(3)$ associated with the links and nodes
of a graph correspondingly, the simplest idea is to take an integral
over these manifolds.
Then the integral over $H\times \Rb$ encodes the functional integration 
over the connection $\SA$ and the integral over $X$ corresponds 
to the integration over the field $\chi$.
This leads to the following expression for the inner product:
\be 
<f_{\Gamma_1},g_{\Gamma_2}>= \int_{[H]^{\#links}} d\mu(h)\, 
 \int_{\Rb^{\#links}} d\fase\, \int_{[X]^{\#nodes}} d\mu(x)\, 
  \overline{f_{\Gamma}(h,\fase,x)}
g_{\Gamma}(h,\fase,x).
\label{sc-pr3}
\ee
(It is implied that both $f_{\Gamma_1}$ and $g_{\Gamma_2}$ were continued
in the trivial way to the common graph $\Gamma=\Gamma_1\cup\Gamma_2$.)
The inner product (\ref{sc-pr3}) is explicitly
Lorentz invariant, since the effect of a Lorentz transformation of the
states can be absorbed into the integration measure over $X$.

Let us calculate the inner product (\ref{sc-pr3}) for the states
(\ref{coef4}). Before performing the integrations, we extend the
integral over $X$ to the whole group $G$. This can be done
since appearing additional matrix elements $D^{\lambda}_{qp}(\hH)$
can be translated to act on $D^{j}_{ba}(h)$ and absorbed into the 
integration over $H$ due to the left-right invariance of the Haar
measure. The remaining integral gives the volume of $H$ which is
normalized to 1.
As a result, we can perform all integrations 
due to the orthogonality of the matrix elements. The result reads
\beq
& <f_{\Gamma_1},g_{\Gamma_2}>= &
\nonumber \\ &
\left(\prod\limits_{i=1}^n\sum\limits_{j_i} 
 \int\limits_{-\infty}^{\infty} d\repr_i \, \repr_i^2  \right)
 \left(\prod\limits_{r=1}^{\nn} \int d\ro(\lambda_r)  
\frac{\sum\limits_{q_r} \sum\limits_{a_{r_k},b_{r_l}}
\left| <\lambda_r, q_r| -{\repr}_{r_k}, (j_{r_k},a_{r_k});
 {\repr}_{r_l}, (j_{r_l},b_{r_l})> \right|^2} 
{N^2_r(\lambda_r; \{\repr_{r_k}, j_{r_k}\},\{ \repr_{r_l}, j_{r_l}\})} 
\sum\limits_{p_r} \right)
%& \nonumber \\ & 
\overline{{\hat f}^{j}_{p}(\lambda; \repr)}
 {\hat g}^{j}_{p}(\lambda; \repr) . &
\label{sc-pr4}
\eeq
Therefore, if we take
\be
  N_r(\lambda_r; \{ \repr_{r_k}, j_{r_k}\},\{ \repr_{r_l}, j_{r_l}\})=
\sqrt{ \sum\limits_{q_r} \sum\limits_{a_{r_k},b_{r_l}}
\left| <\lambda_r, q_r| -{\repr}_{r_k}, (j_{r_k},a_{r_k});
 {\repr}_{r_l}, (j_{r_l},b_{r_l})> \right|^2},
\label{norm}
\ee
the vectors (\ref{basis1}) will form an orthonormal basis.
(Of course, it is implied that the right hand side of Eq. (\ref{norm})
does not vanish, what simply restricts the range of summations and
integrations.)

Notice, that without the variable $\fase$ and the integration over it 
we would remain with two integrals over $\repr_i$ coming from
$f_{\Gamma}$ and $g_{\Gamma}$ correspondingly. 
It means that states with different 
assignments of the Lorentz representations to the links would not be
orthogonal to each other.
Besides, it is interesting to note that we can not add 
the supplementary series to the representations
associated with the links. In this case the states also would not be 
orthogonal despite the integration over $\fase$ since $e^{i\fase\repr}$
becomes real.

The resulting Hilbert space is obtained by completion of the space of 
the generalized cylindrical functions (\ref{coef4})
with respect to the measure induced by Eq. (\ref{sc-pr3}).
This structure is translated to the space of quantum states 
in the ``connection representation''
provided we establish the following correspondence:
\be
\Psi_S \leftrightarrow e^j_p(\lambda;\repr),
\label{basis}
\ee
where $\Psi_S$ is a Lorentz spin network (\ref{spnetA}).
With respect to this structure the Lorentz spin networks form
an orthonormal basis in the Hilbert space of quantum gravity.

We finish this subsection with some comments.
As it was discussed in the end of Sec. II 
there are definite limitations on the use of the identification (\ref{basis}).
However, these limitations do not restrict the physical information
which can be found by use of our construction.
Indeed, consider the calculation of a matrix element of a quantum
operator between two states. Let the states are given in terms
of functions on the homogeneous space. Then to find the matrix
element we should correspond them quantum states expressed in terms of
the Lorentz spin networks (\ref{spnetA}) via
Eq. (\ref{basis}), act by the operator, make the inverse
identification, and calculate the inner product (\ref{sc-pr3})
of two resulting states. Following this procedure we do not arrive to any 
contradictions with the results obtained in other ways.
In particular, the area spectrum is given by Eq. (\ref{areaspL}).

\subsection{Gauge invariant subspace}
   
If we work directly in the gauge invariant subspace,
the situation simplifies drastically.
In this case our state space is realized by functions
$f_{\Gamma}(h,\fase)$ on $\CXG=[H\times\Rb]^n$ invariant under the
transformation (\ref{trans-h}).
The basis is given by
\be
\ve^{j}(\repr)=
\left(\prod\limits_{i=1}^n\sum\limits_{a_i,b_i} 
 D^{j_i}_{b_i a_i}(h_i) e^{i\fase_i\repr_i} \right)
 \prod\limits_{r=1}^{\nn}   
\frac{< \repr_{r_k}, (j_{r_k},a_{r_k})|
 \repr_{r_l}, (j_{r_l},b_{r_l})>}{ 
N_r(\{ \repr_{r_k}, j_{r_k}\}; \{ \repr_{r_l}, j_{r_l}\})},  
\label{basis2}
\ee
where
\be
N_r(\{ \repr_{r_k}, j_{r_k}\}; \{ \repr_{r_l}, j_{r_l}\})=
\sqrt{  \sum\limits_{a_{r_k},b_{r_l}}
\left| <{\repr}_{r_k}, (j_{r_k},a_{r_k})
 |{\repr}_{r_l}, (j_{r_l},b_{r_l})> \right|^2}.
\ee
It is orthonormal with respect to the inner product defined as 
an integral over $\CXG$:
\be 
<f_{\Gamma_1},g_{\Gamma_2}>_{\ph}= \int_{[H]^{\#links}} d\mu(h)\,  
 \int_{\Rb^{\#links}} d\fase\, \overline{f_{\Gamma}(h,\fase)}
g_{\Gamma}(h,\fase).
\label{sc-pr-inv}
\ee
This is the exact result in the sense that there is no problem with
$\tj$-dependence which we encountered considering nongauge invariant 
states (see the discussion in subsection A).

However, it is possible also to describe this
subspace as a part of the space of all nongauge invariant states.
But the description becomes essentially more complicated.
In subsection A we considered the space of square integrable
functions on $\CX$. Therefore the harmonic analysis was relatively simple.
In particular, the measure on the set of unitary irreducible representations
of the Lorentz group was given by the standard Plancherel measure
$\int d\ro(\lambda)=\sum\limits_{n=0}^{\infty}\int\limits_{-\infty}^{\infty}
d\repr\, (n^2+\repr^2)$. 
It vanishes on the supplementary series of representations so that
only the principle series contributes to the decomposition of a square
integrable function \cite{BR}. 

However, it is clear that the gauge invariant states are described by 
not square integrable functions since $\CX$ is a non-compact manifold. 
It is reflected in the fact that
the trivial representation of the Lorentz group
corresponds to $\lambda=(0,\pm 1)$ and does not enter the principle
series of representations. Therefore, it does not appear
in the decomposition of a general state (\ref{coef4}).

Thus, one has to extend the space of functions under consideration.
But in this case it is impossible to introduce a Hilbert space structure
on the extended space which includes the gauge invariant functions on $\CX$.
Besides, the Fourier analysis developed for the space $L^2$
does not work anymore.
A way to overcome these obstacles is to realize our states as functionals
on a dense subset of $L^2(\CX)$. We can choose it to be the space of 
infinitely differentiable functions of compact support $C^{\infty}_0(\CX)$.
There exists an extension of the Fourier analysis on group manifolds
to the case of such generalized functions \cite{Ruhl}.
Therefore, we can apply it to our problem.
However, in this paper we only outline its main steps
and do not enter the mathematical subtleties and details.

The first difference with the previous case happens in Eq. (\ref{coef1}).
Now the integral over $\rep$ is replaced by the integral
along a contour in the complex plane of the parameter $l_1$. 
The position of the contour is defined by the
concrete behaviour of the function ${\breve f}(x)$. In the particular
case of the constant function it consists of 
two circles around $l_1=\pm 1$ \cite{Ruhl}.    
In the similar way the integral $\int d \ro(\lambda)$ in
Eq. (\ref{coef03}) and, consequently, in Eq. (\ref{coef4})
should be properly generalized.

But now we encounter another problem. As it was mentioned, 
the decomposition of the tensor product of two representations 
of the principle series contains only representations of this series 
\cite{Naim}.
Therefore, we have to generalize also the notion of interwiner
to get a nonvanishing result in Eq. (\ref{basis1}).
A general expression for the interwiners can be given in terms of
the integral of group matrix elements \cite{BR}:
\beq
&<\alpha_1; \cdots ;\alpha_k|
\alpha'_1; \cdots ;\alpha'_l>=
N^{-1}_{\beta_1\dots\beta_k,\beta'_1\dots\beta'_l}
\int_G d\mu(g) \prod\limits_{r=1}^k D^{\lambda_r}_{p_r q_r}(g)
\prod\limits_{s=1}^l {D^{\lambda'_s}_{p'_s q'_s}(g^{-1})},&
\label{int}
 \\
&
N_{\beta_1\dots\beta_k,\beta'_1\dots\beta'_l}=
\left( \int_G d\mu(g) \prod\limits_{r=1}^k D^{\lambda_r}_{q_r q_r}(g)
\prod\limits_{s=1}^l {D^{\lambda'_s}_{q'_s q'_s}(g^{-1})}
 \right)^{1/2}, &
\eeq
where we denoted $\alpha=(\lambda,p)$, $\beta=(\lambda,q)$ and
$N_{\beta_1\dots\beta_k,\beta'_1\dots\beta'_l}$ 
is a normalization coefficient. 
(There is no summation over $q_r$ and $q'_s$.)
In fact, the normalization is not essential since it is cancelled in the
combination entering Eq. (\ref{basis1}).

Consider the simplest example of coupling two simple representations and
define its interwiner with the trivial representation. Using
Eq. (\ref{int}) we obtain
\beq
& <\repr_1, p_1| \repr_2, p_2>=<0|-\repr_1, p_1; \repr_2, p_2> &
\nonumber \\
& = N^{-1}_{(\repr_1,q)(\repr_2,q)}\int d\mu(g)
{D^{(0,i\repr_1)}_{p_1q}(g)}
{D^{(0,i\repr_2)}_{p_2q}(g^{-1})}
=N^{-1}_{(\repr_1,q)(\repr_1,q)}
\repr_1^{-2}\delta(\repr_1-\repr_2)\delta_{p_1p_2}. &
\label{int2}
\eeq
Thus, the interwiners become also distributional. 
This is not a problem if we integrate over representations as it is
done in Eq. (\ref{coef4}). However, this may cause that in some cases
the spin networks are not well defined when they are considered on
their own right. For example, this happens 
for a loop with one two-valent node. And, in general, such two-valent
nodes give rise to unphysical infinities due to the $\delta$-function
in Eq. (\ref{int2}). This indicates that either such states
should be regarded only as distributions or the interwiner 
(\ref{int2}) should be redefined.  
In fact, we obtain another infinity due to $N_r$ in the denominator of
Eq. (\ref{basis1}) which is defined by Eq. (\ref{norm}). From the
formal point of view, two infinities exactly cancel each
other. Therefore, it is tempting to redefine the interwiner
(\ref{int2}) replacing $\delta$-function by the Kronecker 
symbol\footnote{The same expression for the two-valent interwiner
should be used in Eq. (\ref{basis2}) where we work directly in the
gauge invariant subspace.}
\be
  <0|-\repr_1, p_1; \repr_2, p_2>=\delta_{\repr_1\repr_2}\delta_{p_1p_2}.
\ee

For higher valent nodes this problem is absent since 
the integral of three matrix elements of the principle series is
always converge and for all representations in the strip $|l_1|\le 1$
matrix elements are bounded functions on the group.
Therefore, in a general case, except the two-valent one,
we define interwiners by the expression (\ref{int}).

Finally, we note that the states which are not described by 
functions of $L^2(\CX)$ remain non-normalizable. 
But it does not mean that they are not physical states.
The fact that the inner product diverges on the gauge invariant states
is just a consequence that we integrate over gauge orbits which have
an infinite volume for the Lorentz group.
The physical inner product should be given by a gauge fixed integral.
Since $\chi$ is a pure gauge variable it is enough to take it to be fixed.
Since the integration over $\chi$ is encoded in the integration over
$X$, the physical inner product can be obtained by dropping this integral.
Thus, it is given by Eq. (\ref{sc-pr-inv}).

\subsection{Relation with SU(2) state space}

It is interesting to see how the SU(2) state space, which the standard
loop quantization is based on, emerges in our approach. 
It is obtained by neglecting the dependence of the functions
$f_{\Gamma}(h,\fase,x)$ on two last arguments $\fase$ and $x$. 
Dropping $x$ can be interpreted as we impose the Lorentz part of the
Gauss constraint.
On the other hand, we saw that neglecting $\fase$ 
is equivalent to working directly in the limit (\ref{WLp}). 
Then our projected Wilson lines are the ordinary SU(2) Wilson lines
(\ref{WL-W2}) with the Ashtekar-Barbero connection (\ref{conAB}). 
The effective Immirzi parameter $\beta_{(j)}$ is
defined by representations (\ref{param}). We stress that it has nothing common 
with the Immirzi parameter $\beta$ appearing in the action.
Moreover, due to the restriction to the simple representations,
$\beta_{(j)}=0$ what is unphysical value for $\beta$.
Also we emphasize that nothing in our construction and results 
depends on $\beta$. It has no a physical meaning in the quantum theory
as it has not in the classical one.

However, whereas the dependence of $x$ does disappear in the gauge invariant
subspace, the dependence of $\fase$ is essential for consistency. 
In other words, we can not neglect it since the limit (\ref{WLp})
does not commute with action of quantum operators.
Therefore, the quantization based only on the SU(2) state space is unavoidably
incorrect. In particular, the area spectrum calculated on such space 
\cite{area,ALarea} is wrong. 
A correct quantization should take into account effects of this
noncommutativity. Adding $\fase$ to the degrees of freedom related
with links is the simplest way to do it. But still, it allows
to achieve the consistency on the level of inner product, but
we do know how to implement quantum operators in the resulting Hilbert
space of functions on the homogeneous space. Therefore,
in our construction it is an auxiliary space and,
considering quantum operators, we have 
to do as it was described in the end of subsection B.

% 
%--------------------------------------------------------------------- 
\section{Conclusion} 
%--------------------------------------------------------------------- 
% 
 
In this paper we continued the construction of covariant loop quantum 
gravity begun in \cite{SA,AV,SAcon}. We investigated the Hilbert space
under the assumption that all area operators are simultaneously
diagonalizable. 
Our results are the following. 1) There is a basis realized by 
Lorentz covariant spin networks which are eigenstates of 
the area operators related to any spacelike surfaces. 2) Only the simple
representations of the Lorentz group of type $(0,i\repr)$
are associated with Wilson lines. 3) Under the conditions described in
the text the elements of the Hilbert space can be identified with functions on 
$[SO(3)\times\Rb]^n\times [SO(3,1)/SO(3)]^{\nn}$. 
The correspondence with spin network states (\ref{spnetA}) is given by
Eq. (\ref{basis}). The gauge invariant
sector is described by functions independent of the last argument.
4) The inner product is defined as an integral over the 
homogeneous space. 5) The noncommutativity problem is solved 
by the restriction to the simple representations only.

One can note a remarkable similarity between these results and 
predictions of Lorentzian spin foam models \cite{sfLor}. 
(For a general review of the spin foam approach, see \cite{Or}.)
The most striking similarity is the appearance of the simple
representations as the only admissible Lorentz representations
associated with links (or faces of a spin foam).
However, the reasons for this restriction are different. In the spin
foam models it is a consequence of the so-called {\it simplicity
condition} \cite{BC,LO} or of the harmonic analysis on
$SO(3,1)/SO(3)$. In our approach the simple representations appear as
the only Lorentz representations into which one can embed any function
on SO(3). 
Another point where two approaches converge is the use of
the homogeneous space $SO(3,1)/SO(3)$.
This allows to hope that it is possible to derive 
a consistent spin foam model from the covariant loop quantum gravity
presented here.

We see that the restriction to the simple representations is essential
for the both approaches. Therefore, it is worth to note an interesting
observation. The eigenvalues of the area operator corresponding to the
representations $(0,i\repr)$ exhaust all spectrum, so that addition
of representations with $n \ne 0$ would lead only to an additional
infinite degeneracy of the eigenvalues (see Eq. (\ref{areaspL})).
This picture is consistent with the so called {\it area
representation},\footnote{Such representation has been suggested by 
D.Vassilevich.} where independent states are labelled by areas
carried by the links. From this point of view there is no reason for
the appearance of the additional degeneracy. 

Let us discuss open questions. The first one is to explain
the appearance of the new variable $\fase$ associated with each link
which seems to be very puzzling. 
This variable has no a classical analogue and appears when we change
the Lorentz Wilson line (\ref{WLp}) by the SO(3) one (\ref{WL-W2}).
We realize that its appearance is related to the noncommutativity of
the limiting procedure used in the definition of the projected Wilson
line (\ref{WLp}) with action of quantum operators 
as explained in the end of Sec. II.
But so far its introduction is simply 
an artificial way to make orthogonal different eigenstates 
of the area operators.
It would be very interesting to understand its origin in more detail.

As it was argued, in general we do not obtain the correct result
if we act by a quantum operator on a function on the homogeneous space 
in the usual way, instead of to consider the action on the
corresponding quantum state before the limit in Eq. (\ref{WLp}) is
taken. Therefore, it would be nice to 
find a realization of the operators directly in the space of such
functions to avoid the indirect procedure described after Eq. (\ref{basis}).
This could be a key for understanding of the nature of the variable $\fase$.

However, there is a large obstacle for the existence of such a
representation. The problem is that on the constructed Hilbert space
the representation of operators fails to be a homomorphism of 
the classical operator algebra. For example, all matrix elements
of the smeared triad operator $\tP_X(\Sigma)$ \cite{AV} vanish.
(Three components vanish due to the presence of the projector
in the commutation relations (\ref{com}) and other three disappear
due to the vanishing of the effective Immirzi parameter $\beta_{(j)}$
for the simple representations.) On the other hand, its square 
corresponds to the square of the area operator and does not vanish.

This fact tells us that, actually, we restricted ourselves to a part 
of a larger state space. This space is spanned by states like (\ref{spnetA})  
but with Wilson lines projected at the end points only. (The projection 
is needed to make the states well defined.
Therefore, it represents a nontrivial result that one can construct
such general well defined Lorentz covariant spin network states.)
Our states with the Wilson lines (\ref{WLp}) 
can be obtained in the limit of an infinite number of the trivial
(two-valent) nodes. 
It may happen that the whole space of the more general states 
is important and can not be neglected.
However, their physical sense is unclear since they are 
eigenstates of only those area operators 
which are defined for surfaces intersecting the graphs
at nodes only.
Besides, for such states we will have troubles with 
the noncommutativity of the connection and with the inner product,
because, as it was emphasized, the Wilson lines projected only at the end 
points do not belong to any group.
Therefore, our hope is that only the limiting subspace considered in
the paper is physically relevant. But the situation is to be clarified.

Since we restricted ourselves to the representations $(0,i\repr)$,
the area spectrum is given by 
\be  
{\cal S}=8\pi \hbar G \sum\limits_{i}
\sqrt{ j_i(j_i+1) + \rho^2_i+1}.  
\label{newas}
\ee
In contrast to the SU(2) result the spectrum (\ref{newas}) is
continuous. What meaning this fact has for the quantum gravity should
be realized more carefully yet. Note only that it still gives
a minimal quanta of area $8\pi \hbar G$ which corresponds to 
$j=\rho=0$. It is interesting that the quanta would not appear 
if we add the supplementary series of representations.
May be its existence can be considered as an indication for a discrete
structure of quantum space.

The related problem is the entropy of a black hole.
Certainly, the derivation of the Bekenstein-Hawking formula found
in the SU(2) case \cite{entropy} 
should be generalized to the present situation.
The continuity of the spectrum seems to be a large obstacle.
At the moment, we do not know how to overcome it. May be only
representations with $\rho=0$ should be taken into account counting 
independent states. This issue deserves a further investigation.

To conclude, we would like to stress that the knowledge of the
structure of the Hilbert space opens a lot of 
possible lines for research in the framework of the covariant loop gravity.
Besides the already discussed problems of the black hole entropy and relation
with spin foam models, one can mention, for instance, the spectrum of
the volume operator. 
Also one can try to generalize
the recently appeared approach to quantum cosmology \cite{boj}.
And may be the most important would be to construct a quantum version
of the Hamiltonian constraint which is polynomial in this case
and, therefore, it is expected to be free of the problems arising
in the SU(2) case \cite{Tim}.

\section*{Acknowledgements} 
The author would like to thank R. Livine, 
V. Lyakhovsky, C. Rovelli,
and D. Vassilevich for stimulating and fruitful discussions.  
The work has been supported in part by European 
network EUROGRID HPRN-CT-1999-00161.

\appendix 
 
% 
%--------------------------------------------------------------------- 
\section{Basics of covariant canonical formalism} 
%--------------------------------------------------------------------- 
% 
 
In this Appendix we list the basic definitions concerning 
the Lorentz covariant canonical formulation. For a more detailed
introduction to it we refer to \cite{SA,AV,SAcon}.

The $3+1$ decomposition of spacetime is chosen to be 
\be
e^0=Ndt+\chi_a E_i^a dx^i, \quad e^a=E^a_idx^i+E^a_iN^idt.
\ee
The multiplets which play the role of canonical variables are
\beq
 & A_i^{ X}=(\omega_i^{0a},\frac12 {\eps^a}_{bc}\omega_i^{bc})
     &{\rm -\ connection\ multiplet}, \nonumber\\
&  \tP_X^{ i}=(\tE^i_a,{\eps_a}^{bc}\tE^i_b\chi_c)
     &{\rm -\ first\ triad\ multiplet}, \label{multHP}\\
&  \tQ_X^{ i}=(-{\eps_a}^{bc}\tE^i_b\chi_c,\tE^i_a)
     &{\rm -\ second\ triad\ multiplet}, \nonumber
 \eeq
where the triad multiplets are related by a numerical matrix 
$\tP^i_X=\Pi_X^Y\tQ^i_Y$. In the formulae the following matrices appear
\be
\Pi^{XY} =\left(
\begin{array}{cc}
0&1 \\ 1&0
\end{array}
\right)\delta_a^b,  \qquad
\R^{XY} =g^{XY}-\frac{1}{\beta}\Pi^{XY}=\left(
\begin{array}{cc}
1& -\frac{1}{\beta} \\
 -\frac{1}{\beta} & -1
\end{array}
\right)\delta_a^b .    \label{mat}
\ee
Also one can introduce the {\it inverse triad multiplets} $\Pt_i^X$ and
$\Qt_i^X$ and {\it projectors} which depend on the field $\chi$ only:
\be
I_{(P)X}^Y= \tP^{ i}_X\Pt_i^{Y}, \qquad
I_{(Q)X}^Y = \tQ^{ i}_X\Qt_i^{Y}.
\ee

If we pass to the shifted connection
\be
\SA_i^X=A_i^X + \frac{1}{2\left(1+\frac{1}{\im^2}\right)}
\R^{X}_{S}I_{(Q)}^{ST}\R_T^Z f^Y_{ZW}\Pt_i^W \G_Y,  \label{spcon}
\ee
where $\G_X$ is the Gauss constraint generating the local Lorentz
transformations,
the Dirac brackets can be given in the simple form:
\be
\{\SA_i^X,\tP_Y^j\}_D= \delta_i^j I_{(P)Y}^X,
\label{com}
\ee
whereas the commutator of two connections is horrible:
\beq
&\{ \int d^3x\, f(x)\SA^X_i(x), \int d^3y\, g(y)\SA^Y_j(y)\}_D=&
\nonumber \\
& \frac{1}{2\left(1+\frac{1}{\beta^2}\right)} R^X_S R^Y_T \int d^3z\, \left[
\left(  K^{ST,l}_{ij} g\p_l f- K^{TS,l}_{ji} f\p_l g \right) +
fg \left( L^{ST}_{ij}-L^{TS}_{ji} \right)
\right], & 
\label{AAcom}
\eeq
where
\beq
K^{ST,l}_{ij}&=& \Pi^{SS'}f_{S'}^{PQ}\left[\tQ^l_P\left(
(\Qt\Qt)_{ij}\IQ^T_Q+\Qt_i^T\Qt_j^Q-\Qt_j^T\Qt_i^Q\right)
+\delta_i^l\IQ^T_Q\Qt_j^P\right]
\nonumber \\
L^{ST}_{ij}&=&
\Pi^{S}_{S'}f^{PQ}_Z \left[ \Qt_j^{S'}\Qt_n^T\Qt_i^Z+
(\Qt\Qt)_{in}\Qt_j^{S'}\IQ^{TZ}+\Qt_i^T\Qt_n^{S'}\Qt_j^Z
\right. \nonumber \\
&& \quad \qquad  - \left.
\Qt_i^T\Qt_j^{S'}\Qt_n^Z+(\Qt\Qt)_{ij}\Qt_n^{S'}\IQ^{TZ}-
\Qt_j^T\Qt_n^{S'}\Qt_i^Z\right] \tQ^l_P\p_l\tQ^n_Q
\nonumber \\
&+& \Pi^{S}_{S'}f^{Q}_{ZP} \left[ \Qt_n^T\Qt_j^P+
(\Qt\Qt)_{jn}\IQ^{TP}-\Qt_j^T\Qt_n^{P}\right]
\IQ^{ZS'}\p_i\tQ^n_Q
\nonumber \\
&+& \Pi_{Z}^{Z'}f^{PQ}_{Z'}\left[
(\Qt\Qt)_{in}\Qt_j^{Z}\IQ^{ST}-(\Qt\Qt)_{in}\Qt_j^T\IQ^{SZ}-
(\Qt\Qt)_{ij}\Qt_n^T\IQ^{SZ}  \right]  \tQ^l_P\p_l\tQ^n_Q
\nonumber \\
&+& \Pi^S_{S'}f^Z_{PQ}\Qt_j^{S'}\Qt_i^Q\IQ^{TP}\p_l\tQ_{Z}^l+
f^Z_{PQ}\Qt_i^P\Qt_j^Q\IQ^T_Z\IQ^{SW}\Pi_W^{W'}\p_l\tQ_{W'}^l.
\eeq
It is implied that repeated 6-dimensional indices are always
contracted with help of the Killing form $g_{XY}$.

%
%---------------------------------------------------------------------
\section{Irreducible representations of the Lorentz group}
%---------------------------------------------------------------------
%

The generators $T_X$ form the $so(3,1)$ algebra with the structure constants
$f_{XY}^Z$:
\be
[T_X, T_Y]=f_{XY}^Z T_Z.
\ee
Let us introduce the notations $T_X=(A_a,-B_a)$ and   
\beq  
&H_+=iB_1-B_2, \qquad H_-=iB_1+B_2, \qquad H_3=iB_3,& \\  
&F_+=iA_1-A_2, \qquad F_-=iA_1+A_2, \qquad F_3=iA_3.&   
\eeq  
These generators commute in the following way:  
\beq  
& [H_+,H_3]=-H_+, \qquad [H_-,H_3]=H_-, \qquad [H_+,H_-]=2H_3, &  
\nonumber \\  
& [H_+,F_+]=[H_-,F_-]=[H_3,F_3]=0, & \nonumber \\  
& [H_+,F_3]=-F_+, \qquad  [H_-,F_3]=F_-, & \\  
& [H_+,F_-]=-[H_-,F_+]=2F_3, & \nonumber \\  
& [F_+,H_3]=-F_+, \qquad [F_-,H_3]=F_-, & \nonumber \\  
& [F_+,F_3]=H_+, \qquad [F_-,F_3]=-H_-, \qquad [F_+,F_-]=-2H_3. &  
\nonumber       
\eeq

An irreducible representation of the Lorentz group is characterized
by two numbers $(l_0,l_1)$, where $l_0\in \Nat/2$ and $l_1\in \Cb$.   
In the space $\CH_{l_0,l_1}$ of this representation one can introduce   
an orthonormal basis   
\be  
\{ \xi_{l,m}\},\qquad m=-l,-l+1,\dots,l-1,l, \quad  
l=l_0,\l_0+1,\dots   
\ee  
such that the generators introduced above act in the 
following way \cite{GMS}:  
\beq  
H_3\xi_{l,m}&=& m\xi_{l,m}, \nonumber \\  
H_+\xi_{l,m}&=& \sqrt{(l+m+1)(l-m)}\xi_{l,m+1}, \label{gauss-rep} \\
H_-\xi_{l,m}&=& \sqrt{(l+m)(l-m+1)}\xi_{l,m-1}, \nonumber \\  
F_3\xi_{l,m}&=& \gamma_{(l)}\sqrt{l^2-m^2}\xi_{l-1,m}+\beta_{(l)}m\xi_{l,m}  
-\gamma_{(l+1)}\sqrt{(l+1)^2-m^2}\xi_{l+1,m}, \nonumber \\  
F_+\xi_{l,m}&=&  
\gamma_{(l)}\sqrt{(l-m)(l-m-1)}\xi_{l-1,m+1}+\beta_{(l)}\sqrt{(l-m)(l+m+1)}  
\xi_{l,m+1} \nonumber \\  
&+& \gamma_{(l+1)}\sqrt{(l+m+1)(l+m+2)}\xi_{l+1,m+1}, \label{boosts-rep} \\
F_-\xi_{l,m}&=&  
-\gamma_{(l)}\sqrt{(l+m)(l+m-1)}\xi_{l-1,m-1}+\beta_{(l)}\sqrt{(l+m)(l-m+1)}  
\xi_{l,m-1} \nonumber \\  
&-& \gamma_{(l+1)}\sqrt{(l-m+1)(l-m+2)}\xi_{l+1,m-1}, \nonumber  
\eeq  
where   
\be   
\beta_{(l)}=-\frac{il_0l_1}{l(l+1)}, \qquad   
\gamma_{(l)}=\frac{i}{l}\sqrt{\frac{(l^2-l_0^2)(l^2-l_1^2)}{4l^2-1}}.  
\ee  
The unitary representations correspond to two cases:   
\beq  
1)& (n,i\rho),  \quad n\in \Nat/2, \ \rho\in \Rb & \quad - \  
principal\ series, \\  
2)& (0,\rho), \quad |\rho|<1, \ \rho\in \Rb & \quad - \ 
supplementary\ series.   
\eeq


\begin{thebibliography}{99} 
 
 

%%%%%%%%%%%%% Review on loop approach 
\bibitem{Rov-dif} 
M.~Gaul and C.~Rovelli, 
{\it Loop Quantum Gravity and the Meaning of Diffeomorphism Invariance} 
Lectures given at the 35th Karpacz Winter School on Theoretical Physics: 
From Cosmology to Quantum Gravity (1999) [gr-qc/9910079]. 
 
 
%%%%%%%%%%%%  Immirzi parameter 
\bibitem{Imir} 
G.~Immirzi, 
%``Quantum gravity and Regge calculus,'' 
Nucl.\ Phys.\ Proc.\ Suppl.\ {\bf 57}, 65 (1997) 
[gr-qc/9701052]. 
 
%%%%%%%%%%% Interpretation of the Immirzi parameter 
\bibitem{Rov-Tim} 
C.~Rovelli and T.~Thiemann, Phys.\ Rev.\ D {\bf 57}, 1009 (1998)
[gr-qc/9705059].


%%%%%%%%%%%%%%%%%% Entropy %%%%%%%%%%%%%%%%%%%%%%%%%%
\bibitem{entropy}
A.~Ashtekar, J.~Baez, A.~Corichi, and K.~Krasnov, Phys.\ Rev.\ Lett.
{\bf 80}, 904 (1998).





%%%%%%%%%%%%%%%%%%%% Connection %%%%%%%%%%%%%%%%%%%%%%%%%%%
\bibitem{SAcon}
S.~Alexandrov,
% Choice of connection in loop quantum gravity,
Phys.\ Rev.\ D {\bf 65}, 024011 (2002)
[gr-qc/0107071].


%%%%%%%%%%% Samuel's idea
\bibitem{Sam}
J.~Samuel,
%``Is Barbero's Hamiltonian formulation a gauge theory of Lorentzian
% gravity?,''
Class.\ Quantum\ Grav.\ {\bf 17}, L141 (2000)
[gr-qc/0005095];
Phys.\ Rev.\ D {\bf 63}, 068501 (2001).



%%%%%%%%%%% Lorentz covariant formalism
\bibitem{SA}
S.~Alexandrov, Class.\ Quantum\ Grav.\ {\bf 17}, 4255 (2000)
[gr-qc/0005085].

%%%%%%%%%%% Area spectrum
\bibitem{AV}
S.~Alexandrov and D.~Vassilevich,
%``Area spectrum in Lorentz covariant loop gravity,''
Phys.\ Rev.\ D {\bf 64}, 044023 (2001) [gr-qc/0103105].



%%%%%%%%%%% Area operator in loop quantum gravity
\bibitem{area}
C.~Rovelli and L.~Smolin, Nucl.\ Phys.\ {\bf B442}, 593 (1995);
J.~Lewandowski, {\it The operators of quantum gravity} lecture at the
workshop on canonical and quantum gravity (Warsaw, 1995);
S.~Frittelli, L.~Lehner, and C.~Rovelli, Class\  Quantum\ Grav.\
{\bf 13}, 2921 (1996).

\bibitem{ALarea}
A.~Ashtekar and J.~Lewandowski, Class.\ Quantum\ Grav. {\bf 14},
A55 (1997) [gr-qc/9602046].



%%%%%%%%%%% Lorentzian spin foam models
\bibitem{sfLor}
A.~Perez and C.~Rovelli,
%``Spin foam model for Lorentzian General Relativity,''
Phys.\ Rev.\ D {\bf 63}, 041501 (2001)
[gr-qc/0009021];
%''3+1 spinfoam model of quantum gravity with spacelike and timelike
% components,''
Phys.\ Rev.\ D {\bf 64}, 064002 (2001) [gr-qc/0011037].


 %%%%%%%%%%%%%%%  Group theory
\bibitem{BR}
A.~O.~Barut and R.~Raczka, {\it Theory of Group Representations
and Applications}
(Polish Scientific Publishers, Warsaw, 1977)



%%%%%%%%%%%%%%%  Measure, scalar product....
\bibitem{cyl}
A.~Ashtekar and J.~Lewandowski, {\it
Representation theory of analytic holonomy
$C^{\star}$ algebras}  in {\it Knots and quantum gravity} ed. J.~Baez
(Oxford: Oxford University Press, 1994);
J.\ Geom.\ Phys.\ {\bf 17}, 191 (1995);
J.\ Math.\ Phys.\ {\bf 36}, 2170 (1995).



%%%%%%%%%%% Real connection
\bibitem{barb}
J.~F.~Barbero, Phys.\ Rev.\ D {\bf 49}, 6935 (1994);
Phys.\ Rev.\ D {\bf 51}, 5507 (1995);
Phys.\ Rev.\ D {\bf 51}, 5498 (1995);
Phys.\ Rev.\ D {\bf 54}, 1492 (1996).


%%%%%%%%%% Path integral for Ashtekar gravity
\bibitem{AVpath}
S.~Yu.~Alexandrov and D.~V.~Vassilevich, Phys.\ Rev.\ D {\bf 58}, 124029
(1998) [gr-qc/9806001].


%%%%%%%%%%%%%%%% Clebsch-Gordoen coefficients %%%%%%%%%%%%%%%%%
\bibitem{Cl-G}
R.~L.~Anderson, R.~Raczka, M.~A.~Rashid, and P.~Winternitz,
J. Math. Phys. {\bf 11}, 1059 (1970).


%%%%%%%%%%%%%%%%% Harmonic analysis %%%%%%%%%%%%%%%%%%%%%%%%%%
\bibitem{Ruhl}
W.~Ruhl,
{\it The Lorentz Group and Harmonic Analysis}
(WA Benjamin Inc, New York, 1970).

%%%%%%%%%%%%%%%%% Decomposition of the tensor product %%%%%%%%%%%%%%%%
\bibitem{Naim}
M.~A.~Naimark, Amer. Msth. Soc. Translations ser. 2, {\bf 36}, 101 (1964).


%%%%%%%%%%%%%%%% Review on spin foams %%%%%%%%%%%%%%%%%%%%%
\bibitem{Or}
D.~Oriti,
% Spacetime geometry from algebra:
% spin foam models for non-perturbative quantum gravity,
Rept. Prog. Phys. {\bf 64}, 1489 (2001) [gr-qc/0106091].

%%%%%%%%%%%%%%% Barrett-Crane model %%%%%%%%%%%%%%%%
\bibitem{BC}
J.~W.~Barrett and L.~Crane, Class. Quantum Grav. {\bf 17}, 3101 (2000)
[gr-qc/9904025].

%%%%%%%%%%%%%%% Livine-Oriti %%%%%%%%%%%%%%%%%%%%
\bibitem{LO}
R.~E.~Livine and D.~Oriti,
Barrett-Crane spin foam model from generalized BF type action for gravity,
gr-qc/0104043.



%%%%%%%%%%%%%% Cosmology %%%%%%%%%%%%%%%%%%%%%%%%%%%
\bibitem{boj}
M.~Bojowald, Class.\ Quant.\ Grav. {\bf 17}, 1489 (2000);
Class.\ Quant.\ Grav. {\bf 17}, 1509 (2000);
Class.\ Quant.\ Grav. {\bf 18}, 1055 (2001);
Class.\ Quant.\ Grav. {\bf 18}, 1071 (2001).

%%%%%%%%%%%%%%% Thiemann's hamiltonian %%%%%%%%%%%%%%%%%%
\bibitem{Tim}
T.~Thiemann, Class.\ Qauntum\ Grav.\ {\bf 15}, 839 (1998) [gr-qc/9606089];
Class.\ Qauntum\ Grav.\ {\bf 15}, 875 (1998) [gr-qc/9606090].


 %%%%%%%%%%%%%%%  Representations of the Lorentz group
\bibitem{GMS}
I.~M.~Gel'fand, R.~A.~Minlos and Z.~Ya.~Shapiro, {\it Representations
of the rotation and Lorentz groups and their applications}
(Pergamon Press, 1963), pages 187-189.

\end{thebibliography}
\end{document}